# Identification of unknown parameters and orders via cuckoo search oriented statistically by differential evolution for non-commensurate fractional order chaotic systems ✩

GAO Fei[a,b,*], FEI Feng-xia[a], XU Qian[a], DENG Yan-fang[a], QI Yi-bo[a]

[a]*Department of Mathematics, School of Science, Wuhan University of Technology, Luoshi Road 122, Wuhan, Hubei,430070, People's Republic of China*
[b]*Signal Processing Group, Department of Electronics and Telecommunications, Norwegian University of Science and Technology, N-7491 Trondheim, Norway*

## Abstract

In this paper, a non-Lyapunov novel approach is proposed to estimate the unknown parameters and orders together for non-commensurate and hyper fractional chaotic systems based on cuckoo search oriented statistically the differential evolution (CSODE). Firstly, a novel Gao's mathematical model is

✩The work was carried out during the tenure of the ERCIM Alain Bensoussan Fellowship Programme, which is supported by the Marie Curie Co-funding of Regional, National and International Programmes (COFUND) of the European Commission. The work is also supported by Supported by Scientific Research Foundation for Returned Scholars from Ministry of Education of China (No. 20111j0032), the HOME Program No. 11044(Help Our Motherland through Elite Intellectual Resources from Overseas) funded by China Association for Science and Technology, the NSFC projects No. 10647141, No.60773210 of China, the Natural Science Foundation No.2009CBD213 of Hubei Province of China, the Fundamental Research Funds for the Central Universities of China, the self–determined and innovative research funds of WUT No. 2012–Ia–035, 2012-Ia-041, 2010–Ia–004, The National Soft Science Research Program 2009GXS1D012 of China, the National Science Foundation for Post–doctoral Scientists of China No. 20080431004.

*Corresponding author
  *Email addresses:* `hgaofei@gmail.com` (GAO Fei ), `1092285218@qq.com` (FEI Feng-xia), `121214020@qq.com` (XU Qian), `986627833@qq.com` (DENG Yan-fang), `manshengyibo@163.com` (QI Yi-bo)
  *URL:* `http://feigao.weebly.com` (GAO Fei )

put and analysed in three sub-models, not only for the unknown orders and parameters' identification but also for systems' reconstruction of fractional chaos systems with time-delays or not. Then the problems of fractional-order chaos' identification are converted into a multiple modal non-negative functions' minimization through a proper translation, which takes fractional-orders and parameters as its particular independent variables. And the objective is to find best combinations of fractional-orders and systematic parameters of fractional order chaotic systems as special independent variables such that the objective function is minimized. Simulations are done to estimate a series of non-commensurate and hyper fractional chaotic systems with the new approaches based on CSODE, the cuckoo search and differential evolution respectively. The experiments' results show that the proposed identification mechanism based on CSODE for fractional-orders and parameters is a successful methods for fractional-order chaotic systems, with the advantages of high precision and robustness.

## 1. Introduction

The applications of fractional differential equations began to appeal to related scientists recently[1–27] in following areas, bifurcation, hyperchaos, proper and improper fractional-order chaos systems and chaos synchronization[1–33].

However, there are some systematic parameters and orders are unknown for the fractional-order chaos systems in controlling and synchronization. It is difficult to identify the parameters in the fractional-order chaotic systems with unknown parameters. Hitherto, there have been at two main approaches in parameters' identification for fractional-order chaos systems.

- Lyapunov way. There have been few results on parameter estimation method of fractional-order chaotic systems based on chaos synchronization[34] and methods for parameter estimation of uncertain fractional order complex networks[35]. However, the design of controller and the updating law of parameter identification is still a tough task with technique and sensitively depends on the considered systems.

- non-Lyapunov way via artificial intelligence methods For examples, such as differential evolution[7] and particle swarm optimization[9]. In which the commensurate fractional order chaos systems and simplest case with one unknown order for normal fractional-order chaos systems are discussed. However, to the best of our knowledge, little work in non-Lyapunov way has been done to the parameters and orders estimation of non-commensurate and hyper fractional-order chaos systems. And there are no general mathematical model has been purposed for all these kinds of identification.

We consider the following fractional-order chaos system with time delays.

$$_{\alpha}D_t^q Y(t) = f(Y(t), Y_0(t), \theta, \tau) \tag{1}$$

where $Y(t) = (y_1(t), y_2(t), ..., y_n(t))^T \in \Re^n$ denotes the state vector. $\theta = (\theta_1, \theta_2, ..., \theta_n)^T$ denotes the original parameters, $\tau = (\tau_1, \tau_2, ..., \tau_n)$ is the time delay. $q = (q_1, q_2, ..., q_n), (0 < q_i < 1, i = 1, 2, ..., n)$ is the fractional derivative orders.

$$f|_{(Y(t),Y_0(t),\theta,\tau)} = (f_1, f_2, ..., f_n)|_{(Y(t),Y_0(t),\theta,\tau)}$$

Normally the function $f$ is known. And the $\theta, q, \tau$ are unknown, then the $\Theta = (\theta_1, \theta_2, ..., \theta_n, q_1, q_2, ..., q_n, \tau_1, \tau_2, ..., \tau_n)$ will be the parameters to be estimated.

Then a correspondent system are constructed as following.

$$_\alpha D_t^q \tilde{Y}(t) = f(\tilde{Y}(t), Y_0(t), \tilde{\theta}, \tilde{\tau}) \tag{2}$$

where $\tilde{Y}(t), \tilde{\theta}, \tilde{q}, \tilde{\tau}$ are the correspondent variables to those in equation (1), and function $f$ are the same. The two systems (1) (2) have the same initial condition $Y_0(t)$.

Then the objective is obtained as following,

$$\Theta^* = \arg\min_{\Theta} F = \arg\min_{\Theta} \sum_t \left\| Y(t) - \tilde{Y}(t) \right\|_2 \tag{3}$$

When some the fractional chaotic differential equations $f = (f_1, f_2, ..., f_n)$ are unknown, how to identify the fractional system? That is,

$$(f_1, f_2, ..., f_n)^* = \arg\min_{(f_1, f_2, ..., f_n)} F \tag{4}$$

Now the problem of parameters estimation (3) become another much more complicated question, fractional-order chaos reconstruction problem[36], to find the forms of fractional order equations as in (4). In Reference [36] a novel non-Lyapunov reconstruction method based on a novel united mathematical model was proposed to reconstruct the unknown equations $(f_1, f_2, ..., f_n)$.

When it comes to the system (1) $_\alpha D_t^q Y(t) = f(Y(t), Y_0(t), \theta)$ with neither $q$ nor $\theta$ are known, the united model are not effective. For the united mathematical model[36], to be identified is only $(f_1, f_2, ..., f_n)$ instead of $q$. That is $_\alpha D_t^q \tilde{Y}(t)$ of the equation (2) are not included. Actually, if the $q$ are taken into consideration in the united model (4), then the basic parameters' setting to be reconstructed in Reference [36] will be basic set $\{\times, \div, +, -\}$ with extra $\{=, D_t^q\}$ etc., and the input variables $\{x_1, x_2, ..., x_n\}$ with $\{Y\}$ extra and etc. Although for the candidates "programs" in Reference [36] the maximum depth of tree is 6, considering the maximum number of nodes per tree is infinite, there will be infinite illegal candidates will be generated. Then, in one hand, the most time-consuming thing for the novel united model(4) is to kill these illegal individuals from the legal individuals. However, these defaults are not solved in Reference [36]. In the other hand, as $q \in D_t^q$ in unknown, it is really difficult to generate an individual with $\{\times, \div, +, -, =, D_t^q\}$, in neither illegal nor legal cases. And up till now, there is no existing way to resolve these defaults. And we can conclude from simulations [36] that the proposed method are much more efficient for the systems with coefficients in $(f_1, f_2, ..., f_n)$ as integers than as improper fractions.

Therefore, to estimate the $q$ of the equation (2) with unknown systematic parameters $\theta$ is still a question to be solved for parameters and orders estimation of non-commensurate and hyper fractional-order chaos systems.

And Cuckoo search (CS) is an relatively new and robust optimization algorithm[37, 38], inspired by the obligate brood parasitism of some cuckoo species by laying their eggs in the nests of other host birds (of other species). The searching performance is mainly based on the Lévy flights mathematically[37–39], which essentially provide a random walk while their random steps are drawn from a Lévy distribution for large steps[37–39]. However, in of CS evolution, the Lévy flights in each main iteration are used twice. It has two results, the CS' searching performance become a little strong but the redundant evaluations for the objective function are generated too. Therefore, some more efforts are need to improve the performance of CS.

To the best of authors' knowledge, there are no methods in non-Lyapunov way for non-commensurate and hyper fractional order chaotic systems' parameters and orders estimation so far. The objective of this work is to present a novel simple but effective approach to estimate the non-commensurate and hyper fractional order chaotic systems in a non-Lyapunov way. And the illustrative reconstruction simulations in different chaos systems system are discussed respectively.

The rest is organized as follows. In Section 2, a general mathematical model not only for fractional chaos parameters identification but also for reconstruction in non-Lyapunov way are newly proposed and analyzed in three sub-models A, B and C. And a simple review was given on non-Lyapunove parameters estimation methods for fractional-order and normal chaos systems. In section 3, a novel methods with proposed united model based on Cuckoo search oriented by differential evolution statistically (CSODE) is proposed. And simulations are done to a series of different non-commensurate and hyper fractional order chaotic systems by a novel method based on CSODE in Section 4. Conclusions are summarized briefly in Section 5.

## 2. Gao's mathematical model for fractional chaos reconstruction and orders estimation in non-Lyapunov way

In this section, a general mathematical model for fractional chaos parameters identification in non-Lyapunov way is proposed. A detail explanation for the general mathematical model will given as following subsections in three aspects, sub-model A, B, and C.

### 2.1. Gao's mathematical model

Now we consider the general forms of fractional order chaos systems (1). To make the system (1) more clear, we take its equivalent form as following system (5).

$$\begin{cases} {_\alpha}D_t^{q_1} y_1(t) = f_1(t, y_1(t), y_1(t-\tau), y_2(t), y_2(t-\tau), ..., y_n(t), y_n(t-\tau)) \\ {_\alpha}D_t^{q_2} y_2(t) = f_2(t, y_1(t), y_1(t-\tau), y_2(t), y_2(t-\tau), ..., y_n(t), y_n(t-\tau)) \\ ... \\ {_\alpha}D_t^{q_n} y_n(t) = f_n(t, y_1(t), y_1(t-\tau), y_2(t), y_2(t-\tau), ..., y_n(t), y_n(t-\tau)) \\ L = (y_1, y_2, ..., y_n) \end{cases} \quad (5)$$

And a correspondent system (6) is constructed as following.

$$\begin{cases} {_\alpha}D_t^{\tilde{q}_1} \tilde{y}_1(t) = \tilde{f}_1(t, \tilde{y}_1(t), \tilde{y}_1(t-\tilde{\tau}), \tilde{y}_2(t), \tilde{y}_2(t-\tilde{\tau}), ..., \tilde{y}_n(t), \tilde{y}_n(t-\tilde{\tau})) \\ {_\alpha}D_t^{\tilde{q}_2} \tilde{y}_2(t) = \tilde{f}_2(t, \tilde{y}_1(t), \tilde{y}_1(t-\tilde{\tau}), \tilde{y}_2(t), \tilde{y}_2(t-\tilde{\tau}), ..., \tilde{y}_n(t), \tilde{y}_n(t-\tilde{\tau})) \\ ... \\ {_\alpha}D_t^{\tilde{q}_n} \tilde{y}_n(t) = \tilde{f}_n(t, \tilde{y}_1(t), \tilde{y}_1(t-\tilde{\tau}), \tilde{y}_2(t), \tilde{y}_2(t-\tilde{\tau}), ..., \tilde{y}_n(t), \tilde{y}_n(t-\tilde{\tau})) \\ \tilde{L} = (\tilde{y}_1, \tilde{y}_2, ..., \tilde{y}_n) \end{cases} \quad (6)$$

To have simple forms, we take $\alpha = 0$.

Then the novel objective function (fitness) equation (7) in this paper come into being from equations (18) and (19) as below.

$$F = \sum_{t=0 \cdot h}^{T \cdot h} \left\| \tilde{L} - L \right\|_2 \qquad (7)$$

Now a novel Gao's mathematical model for fractional chaos reconstruction comes into being as Figure 1 shows, where functions $f = (f_1, f_2, ..., f_n)$, $\tilde{f} = (\tilde{f}_1, \tilde{f}_2, ..., \tilde{f}_n)$, fractional orders $q = (q_1, q_2, ..., q_n)$, $\tilde{q} = (\tilde{q}_1, \tilde{q}_2, ..., \tilde{q}_n)$, time-delays $\tau = (\tau_1, \tau_2, ..., \tau_n)$, $\tilde{\tau} = (\tilde{\tau}_1, \tilde{\tau}_2, ..., \tilde{\tau}_n)$, systematic parameters $\theta = (\theta_1, \theta_2, ..., \theta_n)$, $\tilde{\theta} = (\tilde{\theta}_1, \tilde{\theta}_2, ..., \tilde{\theta}_n)$ respectively.

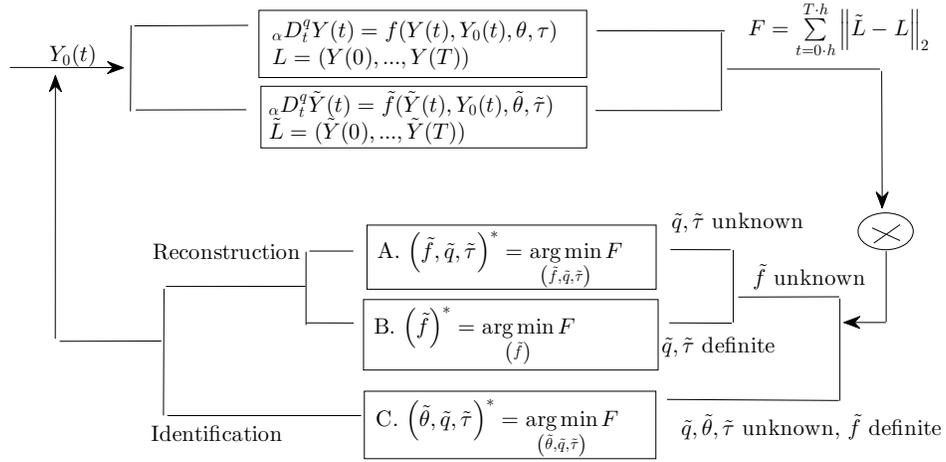

Figure 1: Gao's mathematical model for fractional chaos reconstruction

A detail explanation for the general mathematical model will given as following subsections in three aspects, sub-model A, B, and C.

And the objective function (7) to be optimized can also be any kind of equations (8),(9),(10), (11) to be minimized by artificial intelligence methods as following.

$$F = \sum_{i=1}^{N} \left\| L_i - \tilde{L}_i \right\|_2 \tag{8}$$

$$G = \frac{1}{N} \sum_{i=1}^{N} \left\| L_i - \tilde{L}_i \right\|^2 \tag{9}$$

$$H = \sum_{i=1}^{N} \left\| L_i - \tilde{L}_i \right\|^2 \tag{10}$$

$$W = \frac{1}{N} \sum_{i=1}^{N} \left\| L_i - \tilde{L}_i \right\|_2 \tag{11}$$

2.2. *Mathematical sub-model A.*

It should be noticed here that the independent variables in function (7) in the general model in Figure 1 are not always the parameters and fractional orders. They can be the special variables, for instance, as functions $\tilde{f} = (\tilde{f}_1, \tilde{f}_2, ..., \tilde{f}_n)$, fractional orders $\tilde{q} = (\tilde{q}_1, \tilde{q}_2, ..., \tilde{q}_n)$ and time delays $\tilde{\tau} = (\tilde{\tau}_1, \tilde{\tau}_2, ..., \tilde{\tau}_n)$.

And for the sub-model A, that is

$$\left(\tilde{f}, \tilde{q}, \tilde{\tau}\right)^* = \arg\min_{\left(\tilde{f}, \tilde{q}, \tilde{\tau}\right)} F \tag{12}$$

It can also be written as following.

$$\left((\tilde{f}_1, \tilde{f}_2, ..., \tilde{f}_n), (\tilde{q}_1, \tilde{q}_2, ..., \tilde{q}_n), (\tilde{\tau}_1, \tilde{\tau}_2, ..., \tilde{\tau}_n)\right)^* = \arg\min_{\left((\tilde{f}_1, \tilde{f}_2, ..., \tilde{f}_n), (\tilde{q}_1, \tilde{q}_2, ..., \tilde{q}_n), (\tilde{\tau}_1, \tilde{\tau}_2, ..., \tilde{\tau}_n)\right)} F$$

There exist several definitions of fractional derivatives. Among these, the Grünwald-Letnikov (G-L), Riemann-Liouville (R-L) and the Caputo fractional derivatives are the commonly used[40–45]. And G-L, R-L and Caputo fractional derivatives are equivalent under some conditions[46].

The continuous integro-differential operator[47, 48] is used, and we consider the continuous function $f(t)$. The G-L fractional derivatives are defined as following.

$$_\alpha D_t^q f(t) = \lim_{h \to 0} \frac{1}{h^q} \sum_{j=0}^{\left[\frac{t-\alpha}{h}\right]} (-1)^j \binom{q}{j} f(t - jh) \tag{13}$$

where $[x]$ means the integer part of $x$, $\alpha, t$ are the bounds of operation for $_\alpha D_t^q f(t)$, $q \in \Re$.

We take ideas of a numerical solution method[47, 48] obtained by the relationship (13) derived from the G-L definition to resolve system. That is,

$$_{(k-\frac{L_m}{h})} D_{t_k}^q f(t) \approx \frac{1}{h^q} \sum_{j=0}^{k} (-1)^j \binom{q}{j} f(t_{k-j})$$

where $L_m$ is the memory length, $t_k = kh$, $h$ is the time step of calculation and $(-1)^j \binom{q}{j}$ are binomial coefficients $c_j^{(q)}$, $(j = 0, 1, ...,)$. When for numerical computation, the following are used,

$$c_0^{(q)} = 1, c_j^{(q)} = \left(1 - \frac{1+q}{j}\right) c_{j-1}^{(q)}$$

Then in general, for the simplest case (14) of equation (5) as following.

$$_0 D_t^q f(y(t)) = f(t, y(t), y(t - \tau)) \tag{14}$$

Let $y_{k\tau} = y(t_k - \tau(t_k))$. It can have the approximate value as equation (15), when it used for calculating.

$$y_{k\,\tau} = y_{k+1+[\tau/h]} \frac{\tau/h - [\tau/h]}{h} + y_{k+[\tau/h]} \frac{1 + [\tau/h] - \tau/h}{h} \tag{15}$$

And let $y_k = y(t_k)$, then equation (14) can be expressed as

$$h^{-q} \sum_{j=0}^{k} c_j^{(q)} y(t_{k-j}) = f(t_k, y(t_k), y(t_k - \tau(t_k)))$$

$$y(t_k) = f(t_k, y(t_k), y(t_k - \tau(t_k))) h^q - \sum_{j=v}^{k} c_j^{(q)} y(t_{k-j})$$

$$y_k = f(t_k, y_k, y_{k\tau}) h^q - \sum_{j=v}^{k} c_j^{(q)} y_{k-j} \tag{16}$$

where $v$ in above is defined as

$$v = \begin{cases} k - \frac{L_m}{h}, k > \frac{L_m}{h} \\ 1, k < \frac{L_m}{h} \end{cases}$$

or $v = 1$ for all $k$.

Equation (16) is a implicit nonlinear equation respect to $y_k$. Now we can construct an iteration algorithm to solve $y_k$ as following (17).

$$y_k^{(l+1)} = f\left(t_k, y_k^{(l)}, y_{k\tau}^{(l)}\right) h^q - \sum_{j=v}^{k} c_j^{(q)} y_{k-j} \tag{17}$$

where $l$ is the iteration number. When $\left|y_k^{(l+1)} - y_k^{(l)}\right| < \delta$ ( normally the given error $\delta < 10^{-6}$ ), we consider $y_k = y_k^{(l+1)}$ be the solution of the simplest equation (14). And if the derivative of the $f$ exist and $|f| \leq M$ ($M$ is constant), and $h^q M < 1$, then the iteration (17) converges to a constant as long as the calculus step $h$ is smaller enough.

With the ideas from iteration of equation (17), the systems (5) and (6) are solved as following (18) and (19) respectively,

$$x_i{}_k^{(l+1)} =$$
$$f_i\left(t_{k-1}, t_k, x_1{}_k^{(l)}, x_1{}_{k,\tau}^{(l)}, ..., x_{i-1}{}_k^{(l)}, x_{i-1}{}_{k,\tau}^{(l)}, x_i{}_{k-1}^{(l)}, x_i{}_{k-1,\tau}^{(l)}, ..., x_n{}_{k-1}^{(l)}, x_n{}_{k-1,\tau}^{(l)}\right) h^{q_i}$$
$$- \sum_{j=v}^{k} c_j^{(q)} x_i{}_{k-j}$$
(18)

$$\tilde{x}_i{}_k^{(l+1)} =$$
$$\tilde{f}_i\left(t_{k-1}, t_k, \tilde{x}_1{}_k^{(l)}, \tilde{x}_1{}_{k,\tilde{\tau}}^{(l)}, ..., \tilde{x}_{i-1}{}_k^{(l)}, \tilde{x}_{i-1}{}_{k,\tilde{\tau}}^{(l)}, \tilde{x}_i{}_{k-1}^{(l)}, \tilde{x}_i{}_{k-1,\tilde{\tau}}^{(l)}, ..., \tilde{x}_n{}_{k-1}^{(l)}, \tilde{x}_n{}_{k-1,\tilde{\tau}}^{(l)}\right) h^{\tilde{q}_i}$$
$$- \sum_{j=v}^{k} c_j^{(\tilde{q})} \tilde{x}_i{}_{k-j}$$
(19)

And if $\tau = (0, 0, ..., 0)$, then the systems (5) can be solved as equation (20)[36].

$$x_i{}_k^{(l+1)} = f_i\left(x_1{}_k^{(l)}, ..., x_{i-1}{}_k^{(l)}, x_i{}_{k-1}^{(l)}, ..., x_n{}_{k-1}^{(l)}\right) h^{q_i} - \sum_{j=v}^{k} c_j^{(q_i)} x_i{}_{k-j} \quad (20)$$

To our best of knowledge, there is no work have been done to reconstruct the fractional chaos systems under condition that both $f, q$ and $\tau$ are unknown in sub-model A as equation (12) neither for time-delays free nor with time-delays chaos systems.

2.3. Mathematical sub-model B.

In this sub-model, $f$ are unknown but $\tau$ and $q$ are definite. Then to be estimated is only the fractional differential equations $f$, that is

$$\left(\tilde{f}\right)^* = \arg\min_{(\tilde{f})} F \quad (21)$$

It is should be noticed that there is few method for reconstruction for fractional order chaos systems[36] so far.

However, there are a few results for normal chaos systems, as the special cases of fractional chaos systems. For reconstruction of $f = (f_1, f_2, ..., f_n)$ with the non-Lyapunove methods, they are mainly from symbolic regression through genetic programming(GP)[49–51], and some evolutionary algorithms[52–62].

Considering mathematical sub-model A, we have to say it is really difficult to use the ideas in mathematical sub-model B. Let the input variables are taken as $x, y, z$ and the basic operators set used be $\{+, -, \times, \div, D_t^p, =\}$, where fractional order $p \in [0, 1]$ is uncertain. Now we consider easiest cases that the fractional order differential equation $D_t^{q_1} x = f_1(x, y, z)$ is unknown. Then we will see the individuals as following with the ideas similar to methods for the normal chaos of sub-model B. Figure 2(a) and Figure 2(b) are the normal and correct candidate individuals only for the right part $f_1$ of $dx/dt = f_1(x, y, z)$ of the normal chaos systems.

However, when it comes to fractional order chaos system, the whole fractional order differential equations should be taken into accounts, that is $D_t^{q_1} x = f_1(x, y, z)$ with $q = p \in [0, 1]$ uncertain and $f_1$ unknown. Figure 2(c) shows a correct candidate. And when the evolutions (crossover, mutation and selection) go on, there are some wrong and illegal candidates generated as Figures 2(d), 2(e), 2(f) show. Figure 2(d) is a wrong candidate with $D_t^p(x-y)$. Figure 2(e) is a wrong candidate with $D_t^{p_1}(x-y)$ and $D_t^{p_2}(y)$. Here it should be noticed that random $p_1, p_2 \in [0, 1]$. Figure 2(f) is a wrong candidate with not only $D_t^{p_1}(x - y)$, $D_t^{p_2}(z)$ and $D_t^{p_3}(z)$ but also extra $\{=\}$.

Here it should be noticed that random $p_1, p_2, p_3 \in [0, 1]$.

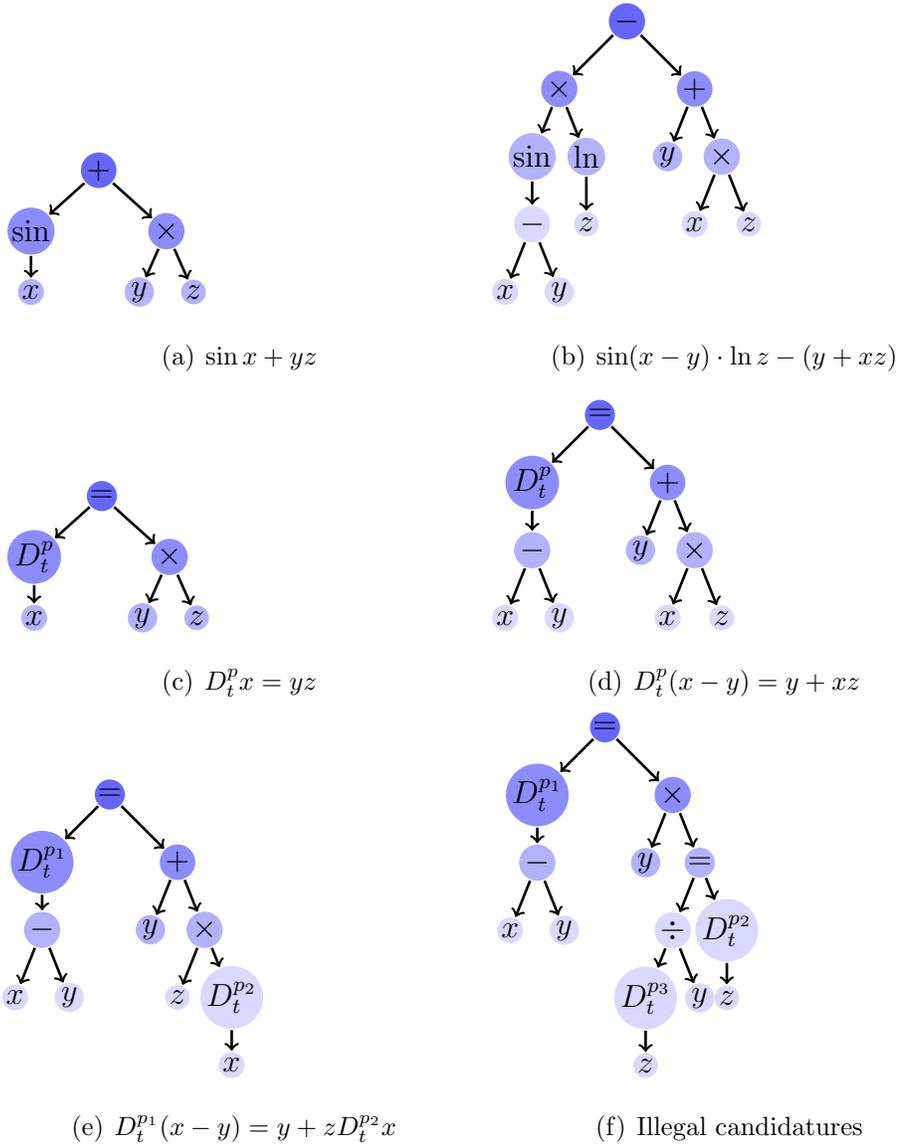

Figure 2: Some examples of the tree structures in GP evolutions

So long as the evolutions (crossover, mutation and selection) go on, these wrong candidate are inevitably existing, although in the genetic program-ming, the tree depth is set to be limited with unlimited leaves. And these kind of wrong individuals will become heavy burden for both the genetic evolution and resolving the fractional order differential equations.

Thus it is not suitable to use the methods based on GP only to reconstruct the fractional order chaos system neither fractional order $q$ nor equations $f_i$ are unknown. However, if it is only considering the unknown equations $f_i$ with definite certain fractional order $q$, these methods will be impressive and efficient as in Reference [36].

2.4. Mathematical sub-model C.

In this sub-model C, $(q_1 \neq q_2 \neq ... \neq q_n)$, systematic parameters and time delays $(\theta_1, \theta_2, ..., \theta_n), \tilde{\tau} = (\tilde{\tau}_1, \tilde{\tau}_2, ..., \tilde{\tau}_n) \in (f_1, f_2, ..., f_n)$ are unknown for non-commensurate and hyper fractional order chaos system.

There some estimation methods have been purposed to identify the unknown parameters and orders for commensurate fractional order chaotic systems. However, to our best of knowledge, no such reconstruction methods have been done for non-commensurate and hyper fractional order chaos system, it is necessary to resolve the following (22) in non-Lyapuno+v way.

$$(\tilde{q}, \tilde{\theta}, \tilde{\tau})^* = \arg \min_{(\tilde{q}, \tilde{\theta}, \tilde{\tau})} F \quad (22)$$

And there exist basic hypotheses in traditional non-Lyapunov estimation methods for fractional order systems[7, 9, 28]. That is, the parameters and fractional orders are partially known or the known data series coincide with definite forms $f = (f_1, f_2, ..., f_n)$ of fractional chaotic differential equations except some uncertain parameters and fractional orders $\Theta = (\theta_1, \theta_2, ..., \theta_n, q_1, q_2, ..., q_n)$.

This is the basic difference between sub-model A,B and C. And for the case when some the chaotic differential equations $f = (f_1, f_2, ..., f_n)$ are unknown, there are some chaos system reconstruction methods. Then the cases[63–75] can be thought as special cases of chaos reconstruction, when the exact forms of chaotic differential equations $f = (f_1, f_2, ..., f_n)$ are available but some parameters unknown.

*2.4.1. Parameters estimation for fractional order chaos systems*

We take the fractional order Lorénz system (23)[3, 8, 24] for example, which is generalized from the first canonical chaotic attractor found in 1963, Lorénz system[76].

$$\begin{cases} {}_\alpha D_t^{q_1} x = \sigma \cdot (y - x); \\ {}_\alpha D_t^{q_2} y = \gamma \cdot x - x \cdot z - y; \\ {}_\alpha D_t^{q_3} z = x \cdot y - b \cdot z. \\ L = (x, y, z) \end{cases} \quad (23)$$

where $q_1, q_2, q_3$ are the fractional orders. When $(q_1, q_2, q_3) = (0.993, 0.993, 0.993)$, $\sigma = 10, \gamma = 28, b = 8/3, \alpha = 0$, intimal point $(0.1, 0.1, 0.1)$ system (23) is chaotic. Generally when the dimension

$$\sum = q_1 + q_2 + q_3 > 2.91$$

for fractional system (23) is chaotic[3, 8, 24].

The form of function (12) can also be as following:

$$p^2 = F(\sigma, \gamma, b, q_1, q_2, q_3) = \sum_{t=0 \cdot h}^{T \cdot h} \left\| \tilde{L} - L \right\|^2 \quad (24)$$

It is noticed that the objective function (24) can be any forms of correspond equations (8),(9),(10), (11).

Then the problems of estimation of parameters for chaotic system are transformed into that of nonlinear function optimization (24). And the smaller $p^2$ is, the better combinations of parameter $(\sigma, \gamma, b, q_1, q_2, q_3)$ is. The independent variables of these functions are $\theta = (\sigma, \gamma, b, q_1, q_2, q_3)$.

And considering the fractional system is very complicated, to simplify the problems, it is reported unknown $q = q_1 = q_2 = q_3, \sigma, \gamma, b$ or case of $\sigma, \gamma, b$ are known and only one $q_i$ are unknown for the similar fractional order chaos systems, such as fractional order Lü system[16, 77] fractional order Chen system[27, 78] fractional Lorénz system[3, 8, 24], discussed in Ref. [7, 9]. The above is the basic idea for the recently proposed methods for fractional chaos system[7, 9].

However, the case $q_1 \neq q_2 \neq q_3$ are not included in the above non-Lyapunov ideas or not fully discussed either for non-commensurate fractional chaos systems.

2.5. The main differences between sub-models A, B and C

Equation (12) is the crucial turning point that changing from the parameters estimation into functions reconstruction and orders estimation, in other words, both fractional order estimation and fractional chaos systems' reconstruction.

It can be concluded that the parameters' estimation of fractional order chaos system[7, 9] is a special case of fractional order chaos reconstruction here as (12). In their researches, the forms of the fractional order differential equations $(f_1, f_2, ..., f_n)$ are known but some parameters $(\theta_1, \theta_2, ..., \theta_n)$ of these equations are unknown, and only one the fractional order and some of these systematic parameters $(\theta_1, \theta_2, ..., \theta_n)$ are estimated[7, 9].

And further, the parameters estimation cases that all $(f_1, f_2, ..., f_n)$ are known but parameters $(\theta_1, \theta_2, ..., \theta_n)$ of these equations are unknown, and the reconstruction case that some of $(f_1, f_2, ..., f_n)$ are unknown, in Section 2.3 for the normal chaos system, are the special cases of fractional order chaos systems' reconstruction (12).

However, it should be emphasized here that, for reconstruction the novel general mathematical model (12) for fractional chaos parameters identification in non-Lyapunov way, with uncertain different fractional order $q$, that is $q_1 \neq q_2 \neq ... \neq q_n \in \{\times, \div, +, -, D_t^{q_1}, D_t^{q_2}, ..., D_t^{q_n}\}$, it is really difficulty to generate a proper candidate from this basic set as shown in Figure 2. Then, it is not easy to reconstruct the fractional order differential equations and identify the fractional orders together. And only the simplest case that with definite $q = q_1 = q_2 = ... = q_n$ are discussed[36].

## 3. Cuckoo search oriented statistically by differential evolution

*3.1. Cuckoo Search*

Cuckoo search (CS) is an optimization algorithm[37, 38], inspired by the obligate brood parasitism of some cuckoo species by laying their eggs in the nests of other host birds (of other species). And some host birds can engage direct conflict with the intruding cuckoos.

CS is based on three idealized rules:

- Only one egg is laid, and is dumped into a randomly chosen nest by each cuckoo at time $t$;

- The best nests with high quality of eggs (candidate solutions) will be copied to the next generation directly;

- The number of available host nests is fixed, and an alien egg will be discovered by a host bird with probability $p_a \in [0,1]$. If so, the host can either throw the egg away or abandon the nest so as to build a completely new nest in a new location.

A Lévy flight is performed for cuckoo $i$ when a new candidates $x^{(t+1)}$ is generated[37–39],

$$x_i^{(t+1)} = x_i^{(t)} + a \oplus \text{Lévy}(\lambda) \qquad (25)$$

where $a > 0$ is the step size which should be related to the scales of the problem of interest. Normally, $a = O(1)$. The product $\oplus$ means entry-wise multiplications. Lévy flights essentially provide a random walk while their random steps are drawn from a Lévy distribution for large steps

$$\text{Lévy} \sim u = t^{-\lambda}, \quad (1 < \lambda \leq 3)$$

which has an infinite variance with an infinite mean, essentially form a random walk process obeying a power-law step-length distribution with a heavy tail[37–39].

Based on above rules and ideas, the basic steps of the CS can be summarised as following pseudo code Algorithm 1.

It should be noticed that in each iteration of Algorithm 1, there are two rounds of evaluation the fitness. ones is after getting a cuckoo by Lévy flights, the other is after abandon the worse nests with probability $p_a$ and building the new nest at the new locations. It is also showed in the original Matlab code in Reference [38].

This might be the reason that CS is efficient. Because CS use Lévy flights twice and evaluate the candidates twice in one generation. However,

**Algorithm 1** The basic steps of the cuckoo search

1: **Basic parameters' setting** Objective function $f(x), x = (x_1, ..., x_d)$, initial population of $n$ host nests $x_i (i = 1, 2, ..., n)$, boundaries for each dimension $x_i = (x_{i,1}, ..., x_{i,d})$ and etc.

2: **while** Termination condition is not satisfied **do**

3:   Get a cuckoo $(i, i = 1, 2, ..., n)$ randomly by Lévy flights (25).

4:   Evaluate its fitness $F_i$;

5:   If $(F_i > F_j)$

6:   Replace $j$ by the new solution

7:   End

8:   Abandon a fraction $(p_a)$ of worse nests.

9:   Build new ones at new locations via Lévy flights (25) $(i, i = 1, 2, ..., n)$.

10:   Keep the best solutions (or nests with quality solutions).

11:   Rank the solutions and find the current best.

12: **end while**

13: **Output** Global optimum $Q_g$

there are one evaluation for the whole population in normal swarm intelligent methods. If we consider the number of evaluating the fitness function by these two evaluations, they might not be economic.

Thus, we can make some modifications here to accelerate the CS as Algorithm 1 by decreasing the evaluation number for the fitness.

*3.2. Differential Evolution Algorithm*

Differential Evolution (DE) algorithm was proposed by Storn [79–82]. DE utilizes $M$ $n$–dimensional vectors, $X_i = (x_{i1}, \cdots, x_{in}) \in S, i = 1, \cdots, M$, as a population for each iteration, called a generation, of the algorithm. For each vector $X_i^{(G)} = (X_{i\,1}^{(G)}, X_{i\,2}^{(G)}, \cdots, X_{i\,n}^{(G)}), i = 1, 2, \cdots, M$, there are three main genetic operator acting[79–82].

To apply the mutation operator, firstly random choose four mutually different individual in the current population $X_{r_1}^{(G)}, X_{r_2}^{(G)}, X_{r_3}^{(G)}, X_{r_4}^{(G)} (r_1 \neq r_2 \neq r_3 \neq r_4 \neq i)$ to compose a differential vector $D^{(G)} = [X_{r_1}^{(G)} - X_{r_2}^{(G)}] + [X_{r_3}^{(G)} - X_{r_4}^{(G)}]$, then combines it with the current best individual $X_{best}^{(G)}$ to get a perturbed vector $V = (V_1, V_2, \cdots, V_n)$ [79, 83] as below:

$$V = X_{best}^{(G)} + CF \times D^{(G)} \qquad (26)$$

where $CF > 0$ is a user-defined real parameter, called mutation constant, which controls the amplification of the difference between two individuals to avoid search stagnation.

Following the crossover phase, the crossover operator is applied on $X_i^{(G)}$. Then a trial vector $U = (U_1, U_2, \cdots, U_n)$ is generated by:

$$U_m = \begin{cases} V_m, & if\ (rand(0,1) < CR)\ or\ (m = k), \\ X_{i\,m}^{(G)}, & if\ (rand(0,1) \geq CR)\ and\ (m \neq k). \end{cases} \quad (27)$$

in the current population[79], where $m = 1, 2, \cdots, n$, the index $k \in \{1, 2, \cdots, n\}$ is randomly chosen, $CR$ is a user-defined crossover constant[79, 83] in the range $[0, 1]$. In other words, the trial vector consists of some of the components of the mutant vector, and at least one of the components of a randomly selected individual of the population.

Then it comes to the replacement phase. To maintain the population size, we have to compare the fitness of $U$ and $X_i^{(G)}$, then choose the better:

$$X_i^{(G+1)} = \begin{cases} U, if\ F(U) < F(X_i^{(G)}), \\ X_i^{(G)}, otherwise. \end{cases} \quad (28)$$

*3.3. Cuckoo search oriented statistically by differential evolution*

Considering the redundant evaluation for the fitness function of CS and the efficiency of DE, we can propose a novel cuckoo search oriented statistically by differential evolution as following Algorithm 2.

In each iteration of Algorithm 2, Lévy flights (25) is used once for each location. And differential evolution operation are used with a probability $p_{De}$ less than 0.2. In this way, the evaluations for the fitness function are reduced nearly 80% compared to the original Algorithm 1.

And $p_{De}$ in Algorithm 2 CSODE should not be too big. Otherwise, it will cause the algorithm 2 be much more like a DE algorithm rather than a cuckoo searcher algorithm. It will be illustrated in the section simulations. Actually, our original idea is let CS oriented not controlled by DE.

**Algorithm 2** Cuckoo search oriented statistically by differential evolution (CSODE)

---
1: **Basic parameters' setting** Objective function $f(x), x = (x_1, ..., x_d)$, initial population of $n$ host nests $x_i (i = 1, 2, ..., n)$, boundaries for each dimension $x_i = (x_{i,1}, ..., x_{i,d})$ and etc.
2: **while** Termination condition is not satisfied **do**
3:     If $p_{De} < 0.2$, generating candidate cuckoo population $(i, i = 1, 2, ..., n)$ randomly from current population by equation (27).
4:     Updating the current cuckoo swarm and the candidate swarm with equation (28).
5:     Abandon a fraction $(p_a)$ of worse nests.
6:     Build new ones at new locations via Lévy flights (25).
7:     Keep the best solutions (or nests with quality solutions).
8:     Rank the solutions and find the current best.
9: **end while**
10: **Output** Global optimum $Q_g$
---

# 4. A novel unknown parameters and orders identification method based on CSODE for non-commensurate fractional order chaos systems

The task of this section is to find a simple but effective approach for unknown $q$ and systematic parameters in equation (22) of for non-commensurate fractional-order chaos based on CSODE in non-Lyapunov way.

*4.1. A novel unknown parameters and orders identification method*

Now we can propose a novel approach for hyper, proper and improper fractional chaos systems. The pseudo-code of the proposed reconstruction is given below.

---
**Algorithm 3** A novel unknown parameters and orders identification method based on differential evolution algorithms for non-commensurate and hyper fractional order chaos systems

---
1: **Basic parameters' setting for Algorithm 2** .
2: **Initialize** Generate the initial population.
3: **while** Termination condition is not satisfied **do**
4:     Algorithm 2 with fitness with Eq. (22).
5:     **Boundary constraints** For each $x_{ik} \in X_i, k = 1, 2, ..., D$, if $x_{i1}$ is beyond the boundary, it is replaced by a random number in the boundary.
6: **end while**
7: **Output** Global optimum $x_{Best}$

---

*4.2. Non-commensurate and hyper fractional order chaos systems*

To test the Algorithm 3, some different well known and widely used non-commensurate and hyper fractional order chaos systems are choose as follow-ing. To have a comparative results, these systems are taken from reference [36].

Example. 1. Here we discuss the non-commensurate fractional Lorénz system (23), so the $(q_1, q_2, q_3) = (0.985, 0.99, 0.99)$ are taken[8].

Example. 2. Fractional order Arneodo's System (29)[47, 84].

$$\begin{cases} {}_0D_t^{q_1} x(t) = y(t); \\ {}_0D_t^{q_2} y(t) = z(t); \\ {}_0D_t^{q_3} z(t) = -\beta_1 x(t) - \beta_2 y(t) - \beta_3 z(t) + \beta_4 x^3(t). \end{cases} \quad (29)$$

when $(\beta_1, \beta_2, \beta_3, \beta_4) = (-5.5, 3.5, 0.8, -1.0)$, $(q_1, q_2, q_3) = (0.97, 0.97, 0.96)$, initial point $(-0.2, 0.5, 0.2)$, Arneodo's System (29) is chaotic.

Example. 3. Fractional order Duffing's system (30)[47].

$$\begin{cases} {}_0D_t^{q_1} x(t) = y(t); \\ {}_0D_t^{q_2} y(t) = x(t) - x^3(t) - \alpha y(t) + \delta \cos(\omega t). \end{cases} \quad (30)$$

when $(a, b, c) = (0.15, 0.3, 1)$, $(q_1, q_2) = (0.9, 1)$, initial point $(0.21, 0.31)$, Duffings system (30) is chaotic.

Example. 4. Fractional order Genesio-Tesi's System (31)[47, 85].

$$\begin{cases} {}_0D_t^{q_1} x(t) = y(t); \\ {}_0D_t^{q_2} y(t) = z(t); \\ {}_0D_t^{q_3} z(t) = -\beta_1 x(t) - \beta_2 y(t) - \beta_3 z(t) + \beta_4 x^2(t). \end{cases} \quad (31)$$

when $(\beta_1, \beta_2, \beta_3, \beta_4) = (1.1, 1.1, 0.45, 1.0)$, $(q_1, q_2, q_3) = (1, 1, 0.95)$, initial point $(-0.1, 0.5, 0.2)$, Genesio-Tesi's System (31) is chaotic.

Example. 5. Fractional order financial System (32)[47, 86] with the exact form of the differential equation $_0D_t^{q_3}z = f_3(x, y, z)$ are unknown.

$$\begin{cases} _0D_t^{q_1}x(t) = z(t) + x(t)(y(t) - a); \\ _0D_t^{q_2}y(t) = 1 - by(t) - x^2(t); \\ _0D_t^{q_3}z(t) = -x(t) - cz(t). \end{cases} \qquad (32)$$

when $(a, b, c) = (1, 0.1, 1)$, $(q_1, q_2, q_3) = (1, 0.95, 0.99)$, initial point $(2, -1, 1)$, financial System (32) is chaotic.

Example. 6. Fractional order Lü system (33)[16, 47].

$$\begin{cases} _0D_t^{q_1}x(t) = a(y(t) - x(t)); \\ _0D_t^{q_2}y(t) = -x(t)z(t) + cy(t); \\ _0D_t^{q_3}z(t) = x(t)y(t) - bz(t). \end{cases} \qquad (33)$$

when $(a, b, c) = (36, 3, 20)$, $(q_1, q_2, q_3) = (0.985, 0.99, 0.98)$, initial point $(0.2, 0.5, 0.3)$, Lü system (33) is chaotic.

Example. 7. Improper fractional order Chen system (34)[27, 47, 78].

$$\begin{cases} _0D_t^{q_1}x(t) = a(y(t) - x(t)); \\ _0D_t^{q_2}y(t) = (d)x(t) - x(t)z(t) + cy(t); \\ _0D_t^{q_3}z(t) = x(t)y(t) - bz(t). \end{cases} \qquad (34)$$

And when when $(a, b, c, d) = (35, 3, 28, -7)$, $(q_1, q_2, q_3) = (1, 1.24, 1.24)$, initial point $(3.123, 1.145, 2.453)$, Chen system (34) is an improper chaotic system[10].

Example. 8. Fractional order Rössler System (35)[12, 47].

$$\begin{cases} _0D_t^{q_1}x(t) = -(y(t) + z(t)); \\ _0D_t^{q_2}y(t) = x(t) + ay(t); \\ _0D_t^{q_3}z(t) = b + z(t)(x(t) - c). \end{cases} \qquad (35)$$

when $(a, b, c) = (0.5, 0.2, 10)$, $(q_1, q_2, q_3) = (0.9, 0.85, 0.95)$, initial point $(0.5, 1.5, 0.1)$, Rössler System (35) is chaotic.

Example. 9. Fractional order Chuas oscillator (36)[87].

$$\begin{cases} {}_0D_t^{q_1}x(t) = \alpha(y(t) - x(t) + \zeta x(t) - W(w)x(t)); \\ {}_0D_t^{q_2}y(t) = x(t) - y(t) + z(t); \\ {}_0D_t^{q_3}z(t) = -\beta y(t) - \gamma z(t); \\ {}_0D_t^{q_4}w(t) = x(t); \end{cases} \quad (36)$$

where

$$W(w) = \begin{cases} a : |w| < 1; \\ b : |w| > 1. \end{cases}$$

when $(\alpha, \beta, \gamma, \zeta, a, b) = (10, 13, 0.1, 1.5, 0.3, 0.8)$, $(q_1, q_2, q_3, q_4) = (0.97, 0.97, 0.97, 0.97)$, initial point $(0.8, 0.05, 0.007, 0.6)$, Chua's oscillator (36) is chaotic.

Example. 10. Hyper fractional order Lorénz System (37)[88].

$$\begin{cases} {}_0D_t^{q_1}x(t) = a(y(t) - x(t)) + w(t); \\ {}_0D_t^{q_2}y(t) = cx(t) - x(t)z(t) - y(t); \\ {}_0D_t^{q_3}z(t) = x(t)y(t) - bz(t); \\ {}_0D_t^{q_4}w(t) = -y(t)z(t) + \gamma w(t); \end{cases} \quad (37)$$

when $(a, b, c, d) = (10, 8/3, 28, -1)$, $(q_1, q_2, q_3, q_4) = (0.96, 0.96, 0.96, 0.96)$, initial point $(0.5, 0.6, 1, 2)$, Hyper fractional order Lorénz System (37) is chaotic.

Example. 11. Hyper fractional order Lü System (38)[89].

$$\begin{cases} {}_0D_t^{q_1}x(t) = a(y(t) - x(t)) + w(t); \\ {}_0D_t^{q_2}y(t) = -x(t)z(t) + cy(t); \\ {}_0D_t^{q_3}z(t) = x(t)y(t) - bz(t); \\ {}_0D_t^{q_4}w(t) = x(t)z(t) + dw(t); \end{cases} \quad (38)$$

when $(a, b, c, d) = (36, 3, 20, 1.3)$, $(q_1, q_2, q_3, q_4) = (0.98, 0.980.98, 0.98)$, initial point $(1, 1, 1, 1)$, Hyper fractional order Lü System (38) is chaotic.

Example. 12. Hyper fractional order Liu System (39)[90].

$$\begin{cases} {}_0D_t^{q_1}x(t) = -ax(t) + by(t)z(t) + z(t); \\ {}_0D_t^{q_2}y(t) = 2.5y(t) - x(t)z(t); \\ {}_0D_t^{q_3}z(t) = x(t)y(t) - cz(t) - 2w(t); \\ {}_0D_t^{q_4}w(t) = -d \cdot x(t). \end{cases} \quad (39)$$

when $(a, b, c, d) = (10, 1, 4, 0.25)$, $(q_1, q_2, q_3, q_4) = (0.9, 0.9, 0.9, 0.9)$, initial point $(2.4, 2.2, 0.8, 0)$, Hyper fractional order Liu System (39) is chaotic.

Example. 13. Hyper fractional order Chen System (40)[91].

$$\begin{cases} {}_0D_t^{q_1}x(t) = -a(y(t) - x(t)) + w(t); \\ {}_0D_t^{q_2}y(t) = dx(t) - x(t)z(t) + cy(t); \\ {}_0D_t^{q_3}z(t) = x(t)y(t) - bz(t); \\ {}_0D_t^{q_4}w(t) = y(t)z(t) + rw(t). \end{cases} \quad (40)$$

when $(a, b, c, d) = (35, 3, 12, 7, 0.5)$, $(q_1, q_2, q_3, q_4) = (0.96, 0.96, 0.96, 0.96)$, initial point $(0.5, 0.6, 1, 2)$, Hyper fractional order Chen System (40) is chaotic.

Example. 14. Hyper fractional order Rössler System (41)[12].

$$\begin{cases} {}_0D_t^{q_1}x(t) = -(y(t) + z(t)); \\ {}_0D_t^{q_2}y(t) = x(t) + ay(t) + w(t); \\ {}_0D_t^{q_3}z(t) = x(t)z(t) + b; \\ {}_0D_t^{q_4}w(t) = -cz(t) + dw(t). \end{cases} \quad (41)$$

when $(a, b, c, d) = (0.32, 3, 0.5, 0.05)$, $(q_1, q_2, q_3, q_4) = (0.95, 0.950.95, 0.95)$, initial point $(-15.5, 9.3, -4, 18.6)$, Hyper fractional order Rössler System (41) is chaotic.

Example. 15. A four-wing fractional order system[92, 93] both incommensurate and hyper chaotic.

$$\begin{cases} D_t^{q_1} x_1 = ax_1 - x_2 x_3 + x_4, \\ D_t^{q_2} x_2 = -bx_2 + x_1 x_3, \\ D_t^{q_3} x_3 = x_1 x_2 - cx_3 + x_1 x_4, \\ D_t^{q_4} x_4 = -x_2, \end{cases} \quad (42)$$

when $(a, b, c) = (8, 40, 49)$, $(q_1, q_2, q_3, q_4) = (1, 0.95 0.9, 0.85)$, initial point $(1, -2, 3, 1)$[92], system (42) is chaotic.

*4.3. Simulations*

For systems to be identified, the parameters of the proposed method are set as following. The parameters of the simulations are fixed: the size of the population was set equal to $M = 40$, generation is set to 500, the default values $CF = 1$, $CR = 0.85, p_{DE} = 0.2$; Table 1 give the detail setting for each system.

Table 2 shows the simulation results of above fractional order chaotic systems. And some simulations are done by single Cuckoo Search (CS) methods.In these cases, all the other parameters for the algorithms are the same as for CSODE. The simulation results are listed in Table 3.

The following figures give a illustration how the self growing evolution process works by DE Algorithm 3. In which, Figures 3,4 ,5,6 7 , 8 , 9,10, 11 show the simulation evolution results of above fractional order chaotic systems with optimization process of objective function's evolution and the parameters and orders uncertain of above fractional order chaotic systems.

From the simulations results of above fractional order chaos system, it can be concluded that the proposed method is efficient. And from above figures,

Table 1: Detail parameters stetting for different systems

| F-O systems | Unknown | Lower boundary | Upper boundary | Step | No. of sam |
|---|---|---|---|---|---|
| Lorénz | $(\sigma, \gamma, b, q_1, q_2, q_3)$ | $5, 20, 0.1, 0.1, 0.1, 0.1$ | $15, 30, 10, 1, 1, 1$ | $0.01$ | $100$ |
| Arneodo | $(\beta_1, \beta_2, \beta_3, \beta_4, q_1, q_2, q_3)$ | $-6, 2, 0.1, -1.5, 0.1, 0.1, 0.1$ | $-5, 5, 1, -0.5, 1, 1, 1$ | $0.005$ | $200$ |
| Duffing | $(a, b, c, q_1, q_2)$ | $0.1, 0.1, 0.1, 0.1, 0.5$ | $1, 1, 2, 1, 1.5$ | $0.0005$ | $500$ |
| Genesio-Tesi | $(\beta_1, \beta_2, \beta_3, \beta_4, q_1, q_2, q_3)$ | $1, 1, 0.1, 0.1, 0.5, 0.5, 0.1$ | $2, 2, 1, 1.5, 1.5, 1.5, 1$ | $0.005$ | $200$ |
| Financial | $(a, b, c, q_1, q_2, q_3)$ | $0.5, 0.01, 0.5, 0.5, 0.1, 0.1$ | $1.5, 1, 1.5, 1.5, 1, 1$ | $0.005$ | $200$ |
| Lü | $(a, b, c, q_1, q_2, q_3)$ | $30, 0.1, 15, 0.1, 0.1, 0.1$ | $40, 10, 25, 1, 1, 1$ | $0.01$ | $100$ |
| Improper Chen | $(a, b, c, d, q_1, q_2, q_3)$ | $30, 0.1, 20, -10, 0.5, 1, 1$ | $40, 10, 30, -0.1, 2, 2, 2$ | $0.01$ | $100$ |
| Rössler | $(a, b, c, q_1, q_2, q_3)$ | $0.1, 0.1, 5, 0.1, 0.1, 0.1$ | $1, 1, 15, 1, 1, 1$ | $0.01$ | $100$ |
| ChuaM | $(\alpha, \beta, \gamma, \zeta, a, b, q_1, q_2, q_3, q_4)$ | $5, 10, 0.1, 0.1, 0.3, 0.1, 0.1, 0.1, 0.1, 0.1$ | $10, 20, 1, 2, 0.3, 1, 1, 1, 1, 1$ | $0.01$ | $100$ |
| Hyper Lorénz | $(a, b, c, d, q_1, q_2, q_3, q_4)$ | $5, 0.1, 20, -2, 0.1, 0.1, 0.1, 0.1$ | $15, 5, 30, -0.1, 1, 1, 1, 1$ | $0.01$ | $100$ |
| Hyper Lü | $(a, b, c, d, q_1, q_2, q_3, q_4)$ | $30, 0.1, 15, 0.1, 0.1, 0.1, 0.1, 0.1$ | $40, 5, 25, 5, 1, 1, 1, 1$ | $0.005$ | $200$ |
| Hyper Liu | $(a, b, c, d, q_1, q_2, q_3, q_4)$ | $5, 0.5, 1, 0.1, 0.1, 0.1, 0.1, 0.1$ | $15, 1.5, 10, 1, 1, 1, 1, 1$ | $0.005$ | $100$ |
| Hyper Chen | $(a, b, c, d, q_1, q_2, q_3, q_4)$ | $30, 0.1, 10, 0.1, 0.1, 0.1, 0.1, 0.1, 0.1$ | $40, 5, 20, 10, 1, 1, 1, 1$ | $0.005$ | $200$ |
| Hyper Rössler | $(a, b, c, d, q_1, q_2, q_3, q_4)$ | $0.1, 0.1, 0.1, 0.01, 0.1, 0.1, 0.1, 0.1$ | $1, 5, 1, 1, 1, 1, 1, 1$ | $0.005$ | $200$ |
| System (42) | $(a, b, c, q_1, q_2, q_3, q_4)$ | $5, 38, 45, 0.5, 0.5, 0.5, 0.5$ | $10, 45, 50, 1, 1, 1, 1$ | $0.005$ | $200$ |

Table 2: Simulation results for different fractional order chaos systems by CSODE

| F-O system | StD | Mean | Min | Max | Success rate[a] | NEOF[b] |
|---|---|---|---|---|---|---|
| Lorénz | 2.2025e-04 | 6.6862e-04 | 3.1179e-06 | 5.5594e-03 | 100% | 24013 |
| Arneodo | 4.7346e-06 | 1.5061e-05 | 3.2217e-07 | 1.4355e-04 | 100% | 24073 |
| Duffing | 5.8073e-09 | 1.4890e-08 | 1.3035e-10 | 1.3548e-07 | 100% | 24016 |
| Genesio-Tesi | 1.3724e-04 | 5.1929e-04 | 7.6119e-07 | 3.8262e-03 | 100% | 24073 |
| Financial | 2.7967e-08 | 2.9039e-08 | 1.5084e-09 | 1.4728e-07 | 100% | 24013 |
| Lü | 2.7950e-04 | 6.5471e-04 | 6.8653e-06 | 5.9367e-03 | 100% | 24038 |
| Improper Chen | 6.5650e-03 | 8.3188e-03 | 1.6981e-04 | 4.3289e-02 | 83% | 24074 |
| Rössler | 1.8666e-08 | 2.1368e-08 | 1.2494e-09 | 1.2324e-07 | 100% | 24013 |
| ChuaM | 1.4857e-04 | 1.4529e-04 | 1.9011e-05 | 1.0265e-03 | 100% | 24073 |
| Hyper Lorénz | 5.1741e-03 | 4.4624e-03 | 5.0835e-04 | 2.4994e-02 | 87% | 24041 |
| Hyper Lü | 1.2904e-02 | 1.2573e-02 | 5.8260e-04 | 5.5329e-02 | 100%[c] | 24041 |
| Hyper Liu | 6.2490e-06 | 4.1206e-06 | 3.3623e-07 | 2.3320e-05 | 100% | 24101 |
| Hyper Chen | 5.4440e-02 | 4.5710e-02 | 6.0972e-03 | 2.7476e-01 | 87% [c] | 24022 |
| Hyper Rössler | 2.0859e-04 | 1.5394e-04 | 3.1938e-05 | 7.9808e-04 | 100% | 24041 |
| System (42) | 4.5612e-01 | 1.8007e+00 | 3.1952e-03 | 1.6640e+01 | 56%[c] | 24024 |

[a] Success means the the solution is less than $1e-2$ in 100 independent simulations.

[b] No. of average evaluation for objective function (NEOF)

[c] Success means the the solution is less than $1e-1$ in 100 independent simulations.

Table 3: Simulation for fractional order systems by single Cuckoo Search

| system | StD | Mean | Min | Max | Success rate[a] | NEOF[b] |
|---|---|---|---|---|---|---|
| Lorénz | 3.8566e-01 | 6.8548e-01 | 1.1083e-01 | 2.2250 | 0% | 40040 |
| Arneodo | 5.2993e-04 | 1.1322e-03 | 2.9726e-04 | 2.7731e-03 | 100% | 40040 |
| Duffing | 2.2560e-04 | 3.8504e-04 | 7.2187e-05 | 1.4730e-03 | 100% | 40040 |
| Genesio-Tesi | 4.0392e-03 | 1.8645e-03 | 1.0365e-03 | 1.0422e-02 | 100% | 40040 |
| Lü | 9.2611e-01 | 1.8303 | 6.0910e-01 | 6.4697 | 0% | 40040 |
| ChuaM | 8.1825e-02 | 4.0495e-02 | 1.5058e-02 | 2.3472e-01 | 0% | 40040 |
| Hyper Lorénz | 9.5917 | 4.4177 | 3.1812 | 2.5456e+01 | 0% | 40040 |
| Hyper Lü | 3.8130 | 7.1730 | 2.5631 | 2.7080 | 0% | 40040 |
| Hyper Chen | 8.8389 | 3.5136 | 2.6274 | 2.4436e+01 | 0% | 40040 |
| system (42) | 9.5876e-02 | 1.4075e-01 | 3.3331e-02 | 5.7142e-01 | 43% | 40040 |

[a] Success means the the solution is less than $1e-1$ in 100 independent simulations.

[b] No. of evaluation for objective function

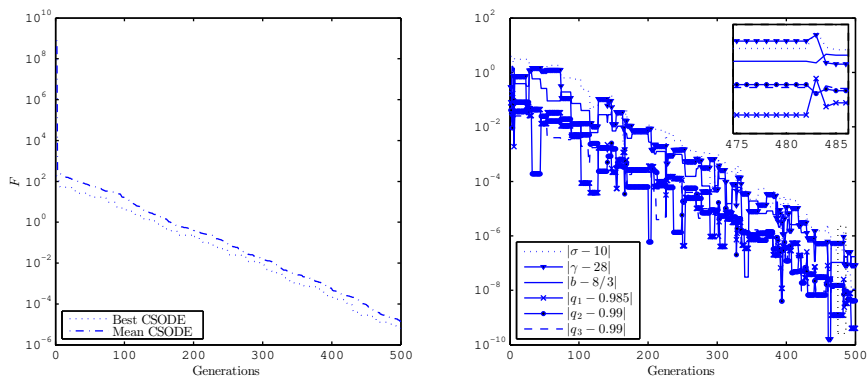

Figure 3: Evolution process for fractional order Lorenz system

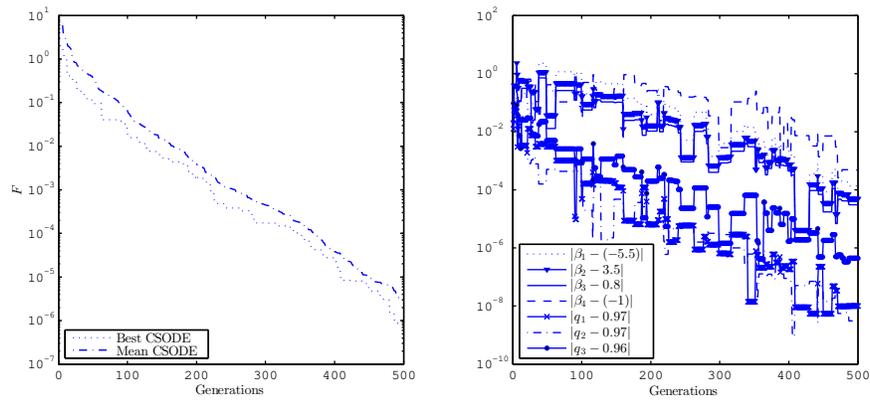

Figure 4: Evolution process for fractional order Arneodo system

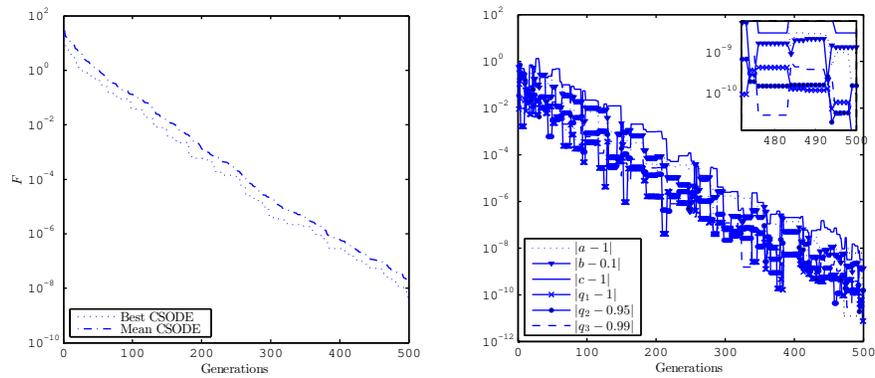

Figure 5: Evolution process for fractional order Financial system

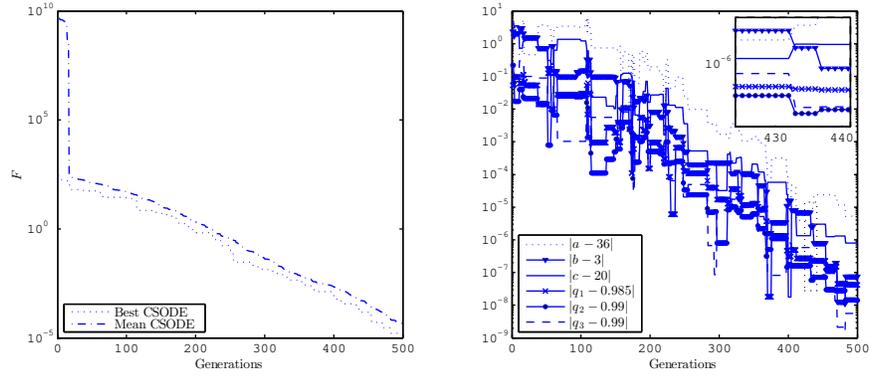

Figure 6: Evolution process for fractional order Lü system

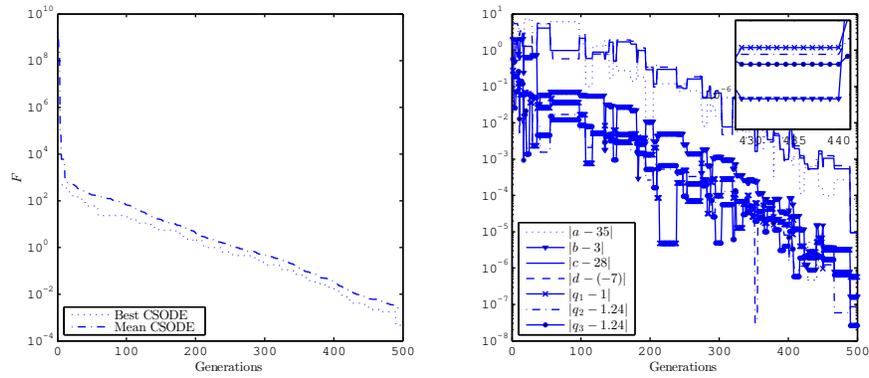

Figure 7: Evolution process for fractional order improper Chen system

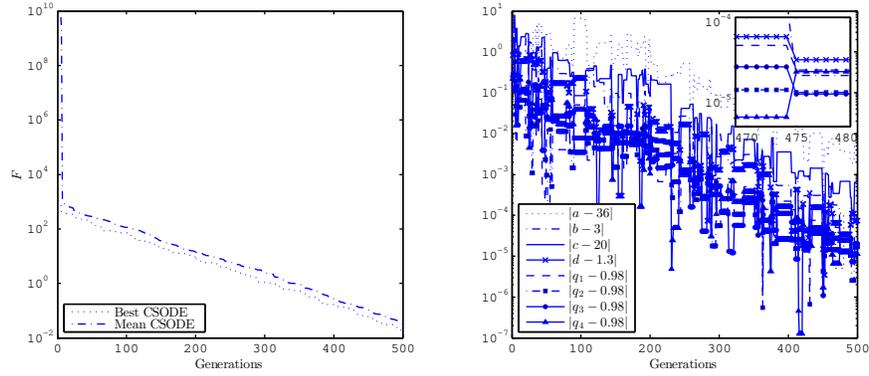

Figure 8: Evolution process for fractional order hyper Lü system

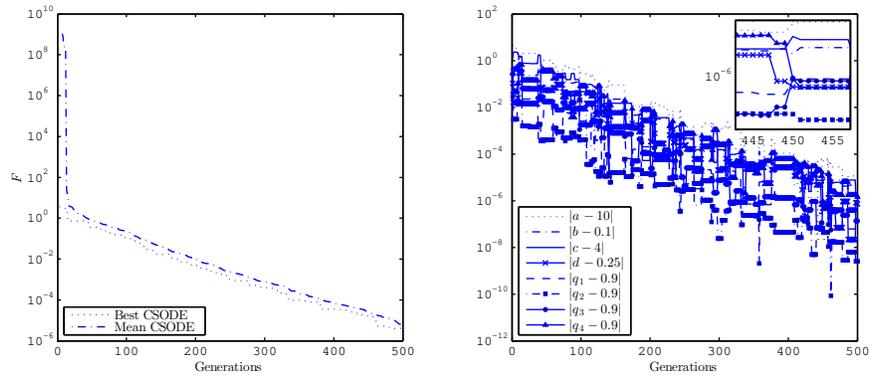

Figure 9: Evolution process for fractional order hyper Liu system

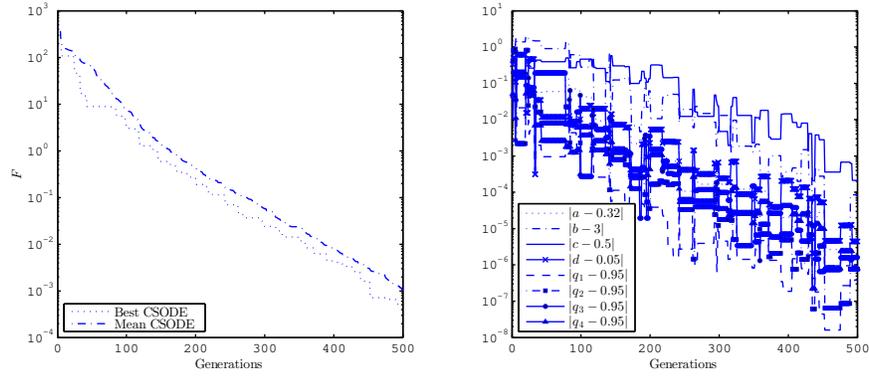

Figure 10: Evolution process for fractional order hyper Rössler system

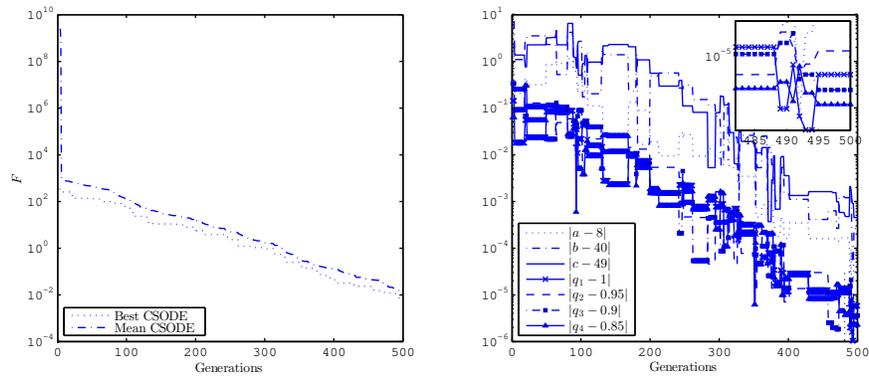

Figure 11: Evolution process for fractional order four wing system (42)

it can be concluded that the estimated systems are self growing under the genetic operations of the proposed methods.

To test the performance of the proposed method Algorithms 3 , some more simulations are done to the four-wing incommensurate hyper fractional order chaotic system (42) in following cases A,B,C,D. In these cases, each with only one condition is changed according to the original setting for system (42). The other parameters for the algorithms are the same as for CSODE. The simulation results are listed in Table 4.

- Case A. Enhancing the defined intervals of the unknown parameters and orders to $[0, 10] \times [30, 45] \times [40, 50] \times [0.1, 1] \times [0.1, 1] \times [0.1, 1] \times [0.1, 1]$ .

- Case B. Reducing the number of samples for computing system (42) from 200 to 100.

- Case C. Increasing the iteration numbers of Algorithms 3 from 500 to 800.

- Case D. Changing the population size of Algorithms 3 from 40 to 80.

Figure 12 show the coresspondent simulation results for system (42).

From results of the Table 2,3, 4 and Figure 12, we can conclude that minimizing the number of samples for computing the system (42) as case B, enhancing the iteration numbers as case C, the population size of Algorithms 3 as case D, will make the Algorithms 3 much more efficient and achieve a much more higher precision. However if the defined intervals of the unknown parameters of system (42) are enhanced, then the results will go to the opposite way. That is the success rate is from 90% to 20% as case A.

Table 4: Simulation results for system (42)

| system (42) | StD | Mean | Min | Max | Success rate | NEOF[b] |
|---|---|---|---|---|---|---|
| Case A. | 2.3271e+02 | 2.2944e+02 | 3.4062e-02 | 5.2168e+02 | 31%[a] | 24000 |
| Case A.[c] | 3.5361e+02 | 4.3992e+02 | 2.9705e-04 | 7.2477e+02 | 39%[a] | 24040 |
| Case B. | 3.3699e-02 | 1.6530e-01 | 1.5989e-05 | 8.3946e-01 | 96% [d] | 23800 |
| Case C. | 6.2779e-03 | 2.4339e-02 | 1.4415e-05 | 1.7882e-01 | 98% [d] | 39360 |
| Case D. | 7.5365e-01 | 2.3014e+00 | 1.1463e-02 | 1.6874e+01 | 41% [d] | 48320 |

[a] Success means the the solution is less than 1 in 100 independent simulations.

[b] No. of evaluation for objective function

[c] By single DE with 600 generations

[d] Success means the the solution is less than $1e-1$ in 100 independent simulations.

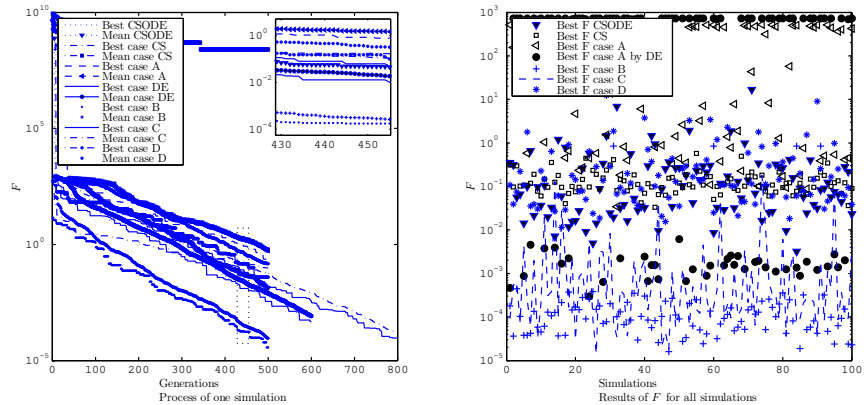

Figure 12: Simulation results for system (42)

If the No. of evaluation for objective function is considered, it is that minimizing the number of samples for computing the system (42) as case B is the best way to achieve higher efficiency and precision.

And according to Table 2,3, the Algorithms 3 based on CSODE is much more better than single cuckoo search.

## 5. Conclusions

The novel Gao's mathematical model in Section 2 is not only for fractional chaos parameters identification but also for reconstruction in non-Lyapunov way with three sub-models A, B and C.

The put method based on CSODE consists of numerical optimization problem with unknown fractional order differential equations to identify the chaotic systems. Simulation results demonstrate the effectiveness and efficiency of the proposed methods with the Gao's mathematical model in Section 2. This is a novel Non–Lyaponov way for fractional order chaos' unknown parameters and orders. The proposed the method solve a the question that the unknown fractional order $q$ are not resolved in reference [36].

Here we have to say that this work is only about the estimation of unknown parameters and orders with the objective function (22) for for non-commensurate and hyper fractional order chaos systems in non-Lyapunov way. It can be concluded that CSODE in Algorithms 3 can be change to other artificial intelligence methods easily. And from the Table 2,3, 4 , we can conclude that Algorithms 3 with CSODE is better than CS. And the the system (42), CSODE is better than DE for the bigger scale of the unknown parameters and orders.

In the future, there are three interesting problems to be studied.

- Neither the fractional orders nor some fractional order equations are unknown. That is, the objective function is chosen as (12) in the novel mathematic model in Section 2. A simple way for this might be the approaches combining the fractional orders and fractional order equations together, that might be both the estimation methods as artificial intelligent methods and the reconstruction methods as in Reference [36] together in some degree.

- Time-delays and systematic parameters are unknown for fractional time-delay chaos systems. The objective function will be selected as the objective function (22) for the time-delay fractional chaotic systems as in Gao's sub-model C , which have special characteristics.

- Cases with noises. Normally, the white noise will be added to the Gao's three sub-models. The similar ideas discussed in Section 2 will be used.

In conclusion, it has to be stated that proposed Algorithms 3 for fractional order chaos systems' identification in a non-Lyapunov way is a promising direction.